\documentclass[10pt,a4paper,oneside,headinclude,footinclude]{scrartcl}
\usepackage{geometry}
\usepackage{graphics}
\usepackage[latin9]{inputenc}
\usepackage[siunitx,europeanresistors,americaninductors]{circuitikz}
\usepackage{tikz}
\usetikzlibrary{shapes.geometric, arrows}
\usepackage{epstopdf}
\usepackage{pgfplots}
\usepackage{amssymb}

\usepackage{siunitx}

\usepackage{setspace}
\title{Cryogenic resonator design for trapped ion experiments in Paul traps}
\author{Matthias F. Brandl\textsuperscript{1}, Philipp Schindler\textsuperscript{1}, Thomas Monz\textsuperscript{1}, and Rainer Blatt\textsuperscript{1,2}}

\begin{document}
\maketitle

{\let\thefootnote\relax\footnotetext{\textsuperscript{1} \textit{Institut f\"ur Experimentalphysik, Universit\"at Innsbruck, Technikerstra{\ss}e 25, A-6020 Innsbruck, Austria}}}
{\let\thefootnote\relax\footnotetext{\textsuperscript{2} \textit{Institut f\"ur Quantenoptik und Quanteninformation der \"Osterreichischen Akademie der Wissenschaften,
Technikerstra{\ss}e 21a, A-6020 Innsbruck, Austria}}}

\begin{abstract}
  Trapping ions in Paul traps requires high radio-frequency
  voltages, which are generated using resonators.  When
  operating traps in a cryogenic environment, an in-vacuum resonator
  showing low loss is crucial to limit the thermal load to the
  cryostat.  In this study, we present a guide for the design and
  production of compact, shielded cryogenic resonators.  We produced
  and characterized three different types of resonators and
  furthermore demonstrate efficient impedance matching of these
  resonators at cryogenic temperatures.
\end{abstract}

\tableofcontents

\section{Introduction}
\label{intro}
Over the last two decades, the application of ion traps has expanded
from mass spectrometry~\cite{Ref:mass-spectrometry} and frequency
standards~\cite{Ref:Ca40-spec,Ref:Sr88-spec} towards engineering of
quantum systems which can be used for quantum
computation~\cite{Ref:Cirac-Zoller,Ref:Kielpinski,Ref:14-qubit} and
quantum
simulation~\cite{Ref:Cirac-Sim,Ref:Quasi-Particle-engineering}.  It is
commonly accepted that large-scale trapped ion quantum information
processors require micrometer-scale ion traps~\cite{Ref:Kielpinski}.
Such traps usually suffer from excessive electric field noise close to
metallic surfaces at room temperature, but at cryogenic temperatures
this noise is strongly
reduced~\cite{Ref:Heating-Rates,Ref:CryoHeatingRates}.  Thus, it seems
natural to move towards cryogenic experimental setups which have the
additional advantage that the ambient pressure is usually a few orders
of magnitude lower than at room temperature. This enables trapping of
longer ion strings due to fewer collisions with background gas.
Furthermore, no bake-out procedure of the vacuum vessel is required,
allowing for rapid trap cycle times. \\
The operation of Paul traps requires high radio-frequency (RF)
voltages, which are usually generated with the aid of the voltage
gain present in RF resonators. For this purpose, helical resonators 
are typically used in the frequency regime up to 50MHz,
whereas for experiments requiring higher drive frequencies coaxial
resonators have been used as
well~\cite{Ref:HelicalResonator,Ref:InductiveCoupling,Ref:CoaxResonator}.
In cryogenic experiments, the resonator needs to fulfill different
criteria than in room temperature experiments where the resonator can
be placed outside the vacuum vessel. In particular, the connections in
a cryostat need to have low thermal conductivity to limit the thermal
load.  Following the Wiedemann-Franz law, this results in a low
electrical conductivity between room-temperature and the cryogenic
parts of the experiment.  Thus, the resonators have to be operated at
the cold-stage.  Moreover, space constraints are stricter in cryogenic
systems which makes helical resonators undesirable as they are
generally bulky\footnote{Helical resonators have been used in
  cryogenic systems~\cite{Ref:CryoResonatorOld}.}.  To minimize the
volume of the
resonator, an RLC-series-resonator can be used ~\cite{Ref:CryoResonatorDavide}. \\
In Section~\ref{sec: general considerations}, we focus on the choice
of the trap drive frequency, the required voltage gain, and voltage
monitoring.  Section~\ref{sec: coil design} covers coil design for
three types of coils, and in the following Section~\ref{sec: matching}, we
discuss impedance matching of the resonator and present an efficient
way to match cryogenic resonators.  Section~\ref{sec: rf shield}
focuses on the design of RF shielding, and finally, we present the results
in Section~\ref{sec: results}.

\section{General considerations}
\label{sec: general considerations}
In trapped ion experiments, RF-voltages of several hundreds of
volts are commonly applied to the trap, which has a simple capacitor as
its electrical circuit equivalent.  Thus, we need to design a resonator
that maximizes the RF-voltage at this capacitor for a given input
power.  In this section, we discuss general considerations in resonator
design which are not limited to a specific resonator type.

\subsection{Choosing the trap drive frequency}
\label{sec: gen considerations - choose trap drive}
At first, we consider the losses in an RLC-resonator and its scaling
with frequency.  From the solution of the equation of motion of a
trapped ion in a Paul trap (Mathieu equation), one obtains the trap
voltages with the stability parameter $q$~\cite{Ref:QuantumSingleIon}, 
which scales as
\begin{equation}
	q = \frac{2 e V}{M r_0^2 \Omega^2} \propto \frac{V}{\Omega^2}
	\label{eq: trap parameter q}
\end{equation}
where $V$ is the amplitude of the RF-voltage at the trap and $\Omega$
is the trap drive (angular) frequency.  Hence, for constant $q$ the
loss power scales as
\begin{equation}
	P \propto V^2 \propto \Omega^4 \hskip 0.5cm .
	\label{eq: losses over f}
\end{equation}
A low trap drive frequency will reduce the losses.  On the other hand,
high secular motion frequencies of the ion in the trap are desirable
for operation.  Higher
secular frequencies require a higher $\Omega$, and thus the desired
secular frequencies set a lower limit for $\Omega$.  In this study, we
aim for a $q$ around 0.25, an axial secular frequency of 1MHz, and
both radial frequencies to be 3.7MHz.  This will require a trap drive
frequency of $\Omega= 2 \pi$ 42.6MHz.  Simulations of our trap show
that we need a drive voltage of about $170V_{\mathrm{rms}}$ to reach the
desired trap frequencies.  In order to limit the thermal load onto the
cryostat from dissipated power in the resonator, this voltage should
be reached with less than 100mW of RF~input power.  This power
corresponds to a consumption of about 1/7l of liquid helium per hour
when operating a wet cryostat at 4.2K~\cite{Ref:Ekin}.

\subsection{Voltage gain of an RLC-resonator}
\label{sec: gen considerations - matching}
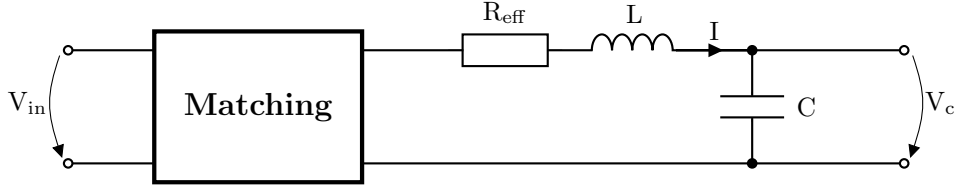
\begin{figure}[ht!]
	\begin{center}
		\begin{circuitikz}[scale=1, transform shape]
			\draw[thick] (0,1.75) to [short, color=black, o-] (5.2,1.75);
			\draw (5,1.75) to [R=$\mathrm{R_{eff}}$] (6.5,1.75);
			\draw[thick] (6.3,1.75) to [short, color=black] (6.9,1.75);
			\draw (6.9,1.75) to [L=$\mathrm{L}$] (8,1.75);
			\draw[thick] (8,1.75) to [short, color=black, i=$\mathrm{I}$] (9,1.75)
				to [short, color=black] (9,1.15);
			\draw (9,1.2) to [C=$\mathrm{C}$] (9,0.8);
			\draw[thick] (9,0.85) to [short, color=black] (9,0.25)
				to [short, color=black, -o] (0,0.25);
			\draw (0,0.35) to [open, v^=$\mathrm{V_{in}}$] (0,1.75) ;
			\draw[thick] (9,0.25) to [short, color=black, *-o] (11,0.25);
			\draw (11,0.35) to [open, v=$\mathrm{V_{c}}$] (11,1.75);
			\draw[thick] (11,1.75) to [short, color=black, o-*] (9,1.75);

			\begin{scope}[every node/.style={draw, ultra thick, fill=white, rectangle, text width=2.5cm, minimum height=2cm, text badly centered}]
					\node at (2.5,1) {\bfseries \large Matching};
			\end{scope}
		\end{circuitikz}
		\caption{An RLC-series-resonator with a matching network used to generate high RF-voltages at the trap}
		\label{fig: general matching}
	\end{center} 
\end{figure}
Fig.~\ref{fig: general matching} shows an RLC-series-resonator
driven through a matching network.  In the following, we will assume
perfect impedance matching with a loss-less matching network.  Thus, we
can set the input power equal to the loss power in the resonator
$P_{\mathrm{input}} = P_{\mathrm{loss,resonator}}$, where $P_{\mathrm{input}} =
\frac{V_{\mathrm{in}}^2}{R_{\mathrm{wave}}}$ with $V_{\mathrm{in}}$ being the input voltage
supplied to the circuit and $R_{\mathrm{wave}}$ the wave impedance of the
connecting cable, commonly $50\Omega$.  We can further write
\begin{equation}
	P_{\mathrm{loss,resonator}} = \left|I\right|^2 \cdot R_{\mathrm{eff}} = \left|\frac{V_{\mathrm{c}}}{\frac{1}{i \Omega C}}\right|^2 \cdot R_{\mathrm{eff}} \hskip 0.5cm,
	\label{eq: loss in the resonator}
\end{equation}
where $I$ is the current in the resonator, $R_{\mathrm{eff}}$ its effective
loss resistance, $\Omega$ its frequency, $V_{\mathrm{c}}$ the voltage at the
capacitor, and $C$ the capacitor representing the trap.  Thus, we
obtain
\begin{equation}
	\frac{V_{\mathrm{in}}^2}{R_{\mathrm{wave}}} = \left|\Omega C V_{\mathrm{in}} G_{\mathrm{V}}\right|^2 \cdot R_{\mathrm{eff}}
	\label{eq: matching 1}
\end{equation}
where $G_{\mathrm{V}}$ is the voltage gain of the circuit, defined as $G_{\mathrm{V}}=\frac{V_{\mathrm{c}}}{V_{\mathrm{in}}}$. \\
The quality factor $Q$ of a resonator is defined as the resonance
frequency $f_{\mathrm{0}}$ divided by the bandwidth $\Delta f$
\begin{equation}
	Q = \frac{f_{\mathrm{0}}}{\Delta f} \hskip 0.5cm.
	\label{eq: quality factor - bandwidth}
\end{equation}
In trapped ion experiments, we are usually not interested in a small
bandwidth around $\Omega$ but rather in a large voltage gain at the
trap drive frequency. It should be noted that the voltage gain and
bandwidth are qualitatively but not necessary quantitatively the same
for different types of resonators. The quantity that is of direct
interest for our application is the voltage gain, which can be derived
from~(\ref{eq: matching 1}) as
\begin{equation}
	G_{\mathrm{V}} = \frac{1}{\Omega \cdot C} \cdot \frac{1}{\sqrt{R_{\mathrm{wave}} \cdot R_{\mathrm{eff}}}} = \sqrt{\frac{Q}{R_{\mathrm{wave}} \cdot \Omega \cdot C}}
	\label{eq: matching 2}
\end{equation}
with the quality
factor of an ideal RLC resonator $Q=\frac{1}{R_{\mathrm{eff}} \cdot \Omega \cdot C}$. 
This indicates that both, the capacitive load,
and the effective resistance of the resonator should be minimized.

\subsection{Pick-up}
\label{sec: gen considerations - pick up}
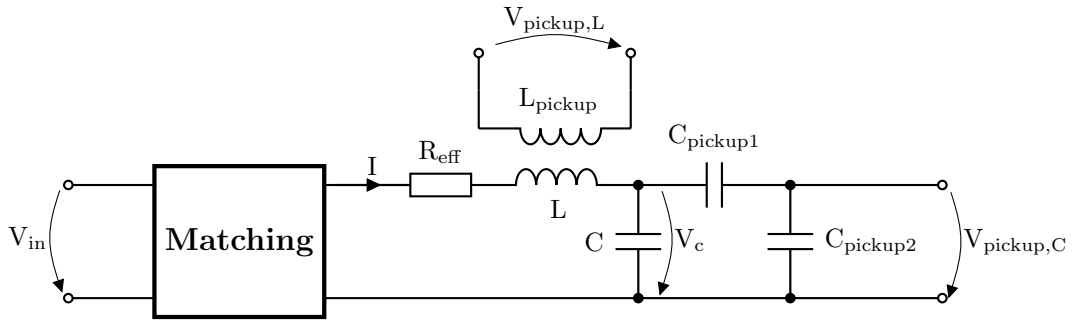
\begin{figure}[ht!]
  \begin{center}
    \begin{circuitikz}[scale=1, transform shape]
      \draw [thick] (0,1.75) to [short, color=black, o-] (3.5,1.75)
				to [short, color=black, i=$\mathrm{I}$] (4.5,1.75);
			\draw (4.4,1.75) to [/tikz/circuitikz/bipoles/length=1cm,R=$\mathrm{R_{eff}}$] (5.4,1.75);
			\draw [thick] (5.3,1.75) to [short, color=black] (5.9,1.75);
			\draw (5.9,1.75) to [L, l_=$\mathrm{L}$] (7,1.75);
			\draw [thick] (7,1.75) to [short, color=black] (7.5,1.75)
				 to [short, color=black] (7.5,1.1);
			\draw (7.5,1.75) to [/tikz/circuitikz/bipoles/length=1cm,C, l_=$\mathrm{C}$] (7.5,0.25);
			\draw [thick] (7.5,0.9) to [short, color=black] (7.5,0.25)
				to [short, color=black, -o] (0,0.25);
			\draw (0,0.35) to [open, v^=$\mathrm{V_{in}}$] (0,1.75) ;
			\draw (7.5,1.75) 
				to [/tikz/circuitikz/bipoles/length=1cm,C=$\mathrm{C_{pickup1}}$, color=black, *-] (9.5,1.75)
				to [/tikz/circuitikz/bipoles/length=1cm,C=$\mathrm{C_{pickup2}}$] (9.5,0.25)
				to [short, color=black, -*] (7.5,0.25);
			\draw [thick] (7.5,1.75) to [short, color=black, *-] (8.4,1.75);
			\draw (9.5,0.25) to [short, color=black, *-o] (11.5,0.25)
				to [open, v=$\mathrm{V_{pickup,C}}$] (11.5,1.75)
				to [short, color=black, o-*] (9.5,1.75);
			\draw [thick] (8.6,1.75) to [short, color=black, -o] (11.5,1.75);
			\draw [thick] (11.5,0.25) to [short, color=black, o-*] (7.5,0.25);
			\draw [thick] (9.5,0.25) to [short, color=black, *-] (9.5,0.9);
			\draw [thick] (9.5,1.75) to [short, color=black, *-] (9.5,1.1);
			\draw [thick] (7.4,3) to [short, color=black] (7.4,2.5)
				 to [short, color=black] (7,2.5);
			\draw (7,2.5) to [L, l_=$\mathrm{L_{pickup}}$] (5.9,2.5);
			\draw [thick](5.9,2.5) to [short, color=black] (5.4,2.5)
				 to [short, color=black] (5.4,3);
			\draw [thick] (7.4,3) to [short, color=black, -o] (7.4,3.5);
			\draw (7.4,3.5) to [open, v=$\mathrm{V_{pickup,L}}$] (5.4,3.5);
			\draw [thick] (5.4,3.5) to [short, color=black, o-] (5.4,3);
			\draw (7.7, 0.3) to [open, v=$\mathrm{V_{c}}$] (7.7,1.75);
			\begin{scope}[every node/.style={draw, ultra thick, fill=white, rectangle, text width=2cm, minimum height=2cm, text badly centered}]
					\node at (2.25,1) {\bfseries \large Matching};
			\end{scope}
		\end{circuitikz}
    \caption{The circuit of an RLC-series-resonator with an inductive and a capacitive pick up. }
	\label{fig: general pickup}
  \end{center}
\end{figure}
Voltage pick-ups are not required for the operation of the trap, but
are useful to measure the voltage on the trap and are required to
actively stabilize the RF voltage on the trap.  In this section, we
discuss inductive and capacitive pick-up, as shown in 
Fig.~\ref{fig: general pickup}. \\
A capacitive pick-up consists of a voltage divider parallel to the
trap, where $C_{\mathrm{pickup2}} \gg C_{\mathrm{pickup1}}$, which leads to the pick-up
voltage
\begin{equation}
	V_{\mathrm{pickup,C}} = \frac{C_{\mathrm{pickup1}}}{C_{\mathrm{pickup1}}+C_{\mathrm{pickup2}}} \cdot V_{\mathrm{c}} \hskip 1cm .
	\label{eq: cap voltage divider}
\end{equation}
$C_{\mathrm{pickup1}}$ has to be small compared to the trap capacitance $C$, or
the capacitive load of the RLC-resonator will increase significantly,
reducing $G_{\mathrm{V}}$ as shown in~(\ref{eq: matching 2}).
Typical values for $C_{\mathrm{pickup2}}$ are several hundred times the value of $C_{\mathrm{pickup1}}$. \\
In an ideal RLC-series-resonator, the voltage on the coil is the same
as the voltage at the capacitor but with a \ang{180} phase shift.  We
can monitor the voltage at the coil as a signal proportional to the
voltage at the trap since we are only interested in the amplitude of
the trap voltage.  Here, $L$ and $L_{\mathrm{pickup}}$ are coupled, and the
pick-up voltage can be estimated with the derivations from
ref.~\cite{Ref:InductiveCoupling}. \\
In order to maintain low losses, the pick-up should not add
significant losses in the resonator.  If one uses a coaxial cable with
a wave impedance $R_{\mathrm{wave}}$ to monitor the pick-up signal, a resistor
$R_{\mathrm{wave}}$ is used as a termination to avoid reflections.  The losses
are then $V_{\mathrm{pickup}}^2/R_{\mathrm{wave}}$ which need to be
small compared to the losses in the resonator $P$. \\
In general, we recommend using an inductive pick-up, because it does
not increase the capacitive load, which would reduce the voltage gain.
However for experiments, which frequently test different traps, a
capacitive pick-up may be preferable since the more accurate ratio
between pick-up and applied voltage facilitates estimating trapping
parameters.

\section{Coil design}
\label{sec: coil design}
Surface traps for cryogenic environments typically use a high-quality
dielectric carrier material and thus the losses in the resonator are
usually dominated by the coil.  In this study, we demonstrate the
design and production of compact and low-loss coils that should allow
us to drive our resonator under the conditions stated in
Section~\ref{sec: gen considerations - choose trap drive}.  Coils
produced with machines are desirable because the production process is
reproducible.  Furthermore, we prefer toroid coils as they guide the
magnetic field in their center, making them less sensitive to their
environment.  In our application, ferromagnetic core materials are
undesirable, because they would induce spatial inhomogeneities of the
magnetic field near the trap.  We estimate the load capacitance of our
resonator to be 10pF, which at a given resonance frequency of 42.6MHz
yields a required inductance of 1.4$\mu$H.

\subsection{PCB Coils}
\label{sec: PCB coils}
Kamby and coworkers~\cite{Ref:PCBCoils} demonstrated the integration
of toroidal RF-inductors into printed circuit boards (PCB).  These
coils can be fully produced by machines and in this study we will
refer to these inductors as PCB coils. \\
Ref.~\cite{Ref:PCBCoils} provides specific formulas for the
inductance $L$, the resistance of one segment $R_{\mathrm{seg}}$, and the
resistance of one via $R_{\mathrm{via}}$ of PCB coils.  In order to minimize
the losses of the coil we want to minimize the resistance of one
winding while maintaining a constant cross-section to keep the
inductance constant.  The resistance of one winding is defined by
\begin{eqnarray}
	2 R_{\mathrm{seg}} + 2 R_{\mathrm{via}} & = 2 \frac{R_{\mathrm{seg}}}{r_{\mathrm{o}} - 
		r_{\mathrm{i}}} \left(r_{\mathrm{o}} - r_{\mathrm{i}}\right) + 2 \frac{R_\mathrm{{via}}}{h} h = \nonumber \\ 
	& = 2 R_{\mathrm{seg}}^\prime \left(r_{\mathrm{o}} - r_{\mathrm{i}}\right) + 2 R_{\mathrm{via}}^\prime h \hskip 0.5cm ,
	\label{eq: pcb coils minimize resistance}
\end{eqnarray}
where $r_{\mathrm{o}}$ and $r_{\mathrm{i}}$ are the outer and the inner radius of the PCB
coil, $h$ is the thickness of the PCB, and $R_{\mathrm{seg}}^\prime$ and
$R_{\mathrm{via}}^\prime$ are the average length-dependent resistances of a
segment and a via.  $R_{\mathrm{seg}}^\prime$ and $R_{\mathrm{via}}^\prime$ depend on
the exact geometry, but usually vias have similar cross-sections as
the segments making their length-dependent resistances approximately
equal.  That means, in order to minimize~(\ref{eq: pcb coils minimize
  resistance}) and to keep the cross section of the coil constant,
$\left(r_{\mathrm{o}} - r_{\mathrm{i}}\right)$ and $h$ should be similar in size. \\
We chose Rogers 4350B as the substrate material, because it has a
small loss tangent, a similar thermal expansion coefficient as copper,
and is ultra-high vacuum compatible.  The thickest available Rogers
4350B had 3.1mm thickness, which is unfortunately too thin to minimize
the resistance for a given cross-section. \\
\begin{figure}[!htb]
	\begin{center}
		\includegraphics[scale=.5]{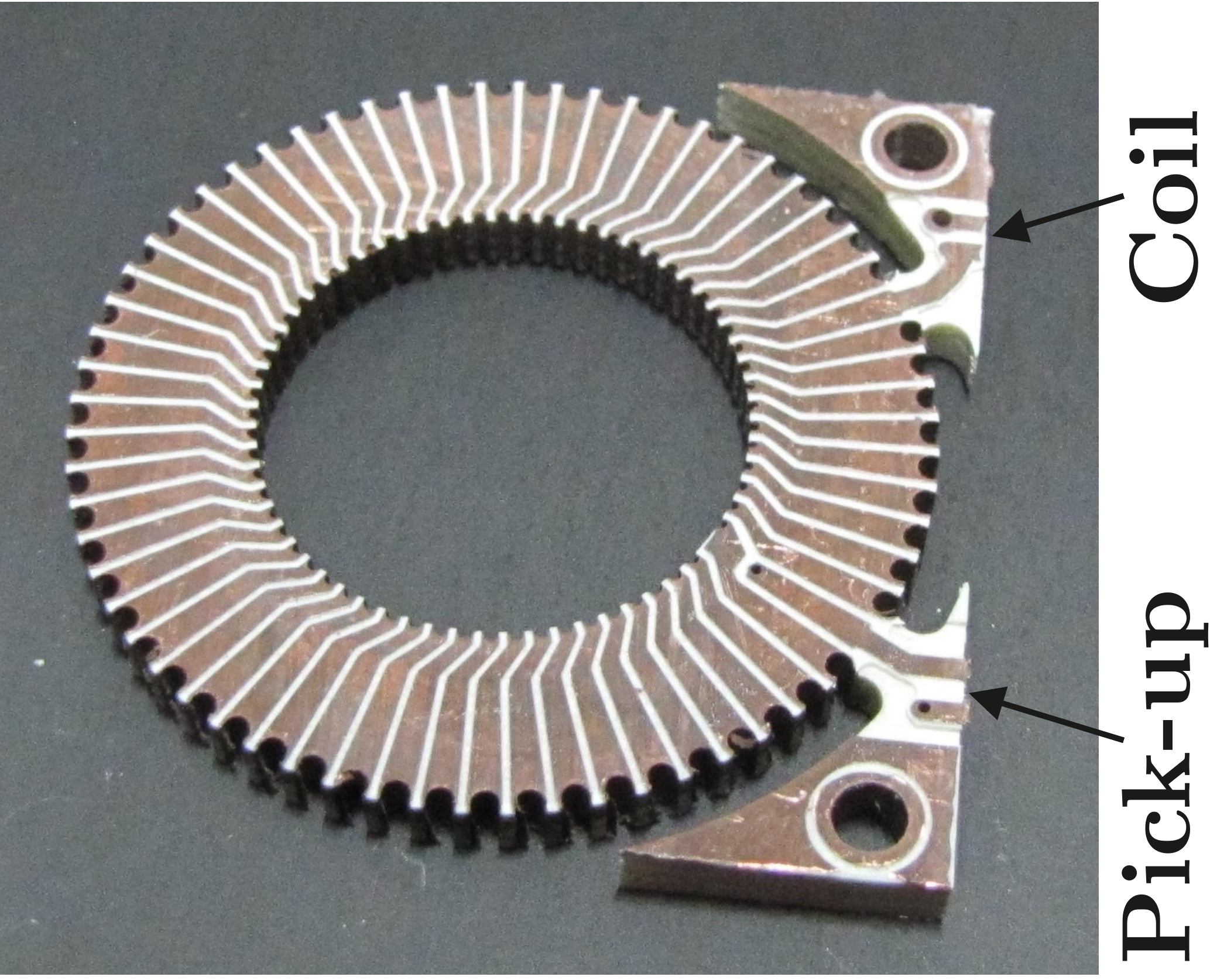}
		\caption{PCB coil produced in copper with a diameter of 35mm and 64 windings. Additionally, there is a pick-up coil
		  included in the design with just one winding.}
		\label{fig: PCB coil}
	\end{center}
\end{figure}
We produced this type of coil in-house and during multiple iterations,
we varied the size and the number of vias.  The lowest losses for
our production process could be achieved with the design depicted in
Fig.~\ref{fig: PCB coil}.  The PCB coil has an outer diameter of 35mm
and an inner diameter 20.4mm with a substrate thickness of 3.1mm.  Its
64 windings result in an inductance of 1.4$\mu$H.  We added one
winding which is galvanically isolated from the rest of the coil.  This
winding acts as an inductive pick-up and should thus be
as reproducible as the entire coil. 

\subsection{Wire Coils}
\label{sec: wire coils}
\begin{figure}[!htb]
	\begin{center}
		\includegraphics[scale=.3]{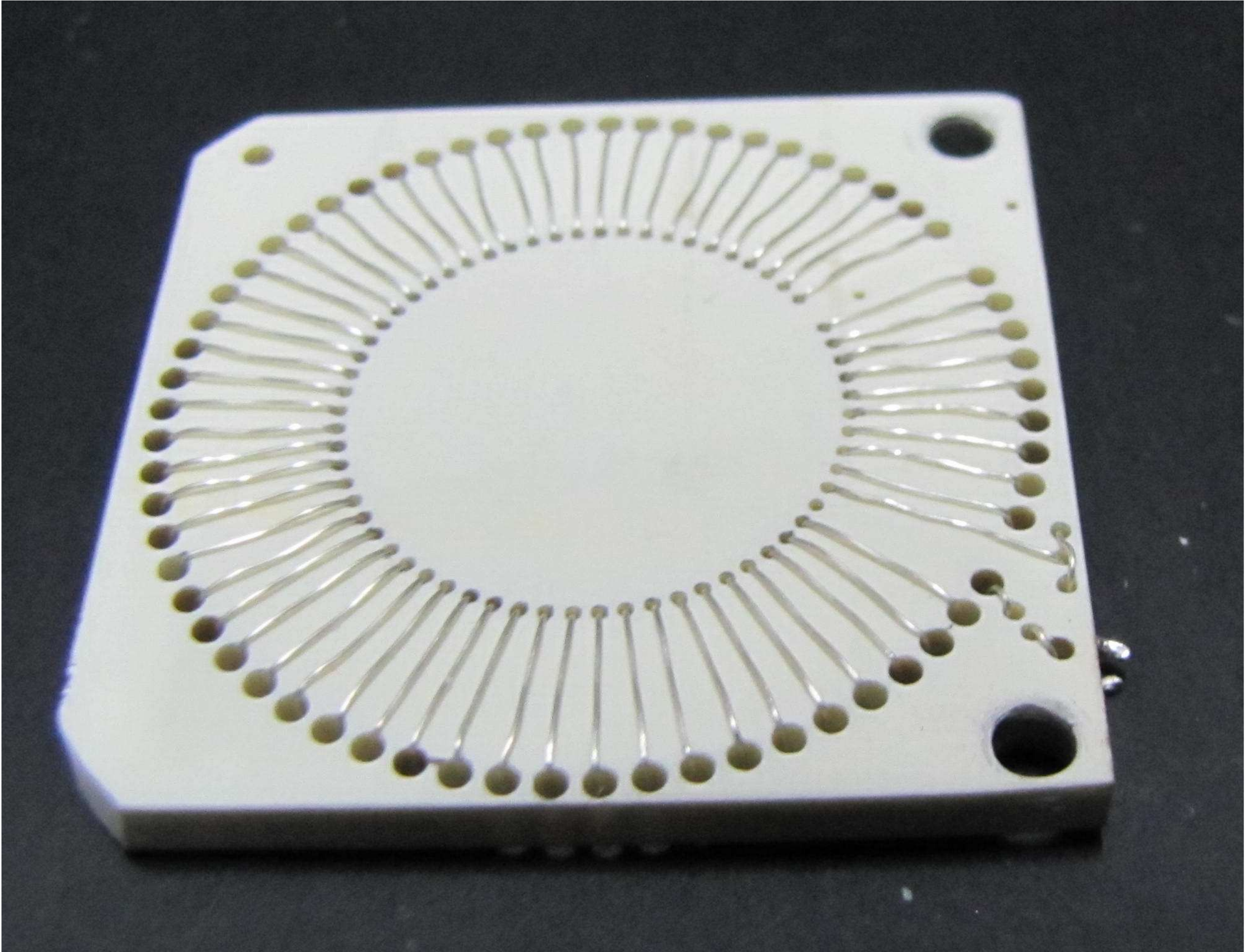}
		\caption{A wire coil made out of 0.4mm thick silver-plated copper wire on a 3.1mm thick Rogers 4350B core.}
		\label{fig: wire coil}
	\end{center}
\end{figure}

For the second type of coils, wire coils, we wound a single wire around
a rigid structure leading to a similar geometry as the PCB coils.  Such
a coil is less reproducible than a PCB coil since the wire length will 
vary for each coil. Hence, its inductance and its resistance will vary
as well.  A coil with the same geometry
as for the PCB coils above, made out of
silver-plated copper on a Rogers 4350B substrate, is shown in 
Fig.~\ref{fig: wire coil}. \\
In wire coils, the cross-section of the core of a wire coil is bigger
than that of a PCB coil, because the wire bends are less sharp.  Hence,
we expect the inductance of a wire coil to be higher than the one of a
PCB coil with the same nominal geometry.  The 0.4mm thick
silver-plated copper wire, which we used for the wire coil, had a
cross-section similar to the average cross-section of a PCB
coil.  Therefore, we expect similar losses in both types of coils with
the same geometries.
These considerations allow us to characterize the production processes 
of the PCB coils by comparing the two types of coils.
Higher losses in the PCB coils are an indication that the production 
of the PCB coils can be improved. \\
Since we do not need traces on the substrate, it is possible use any
machinable material with a low loss tangent.  We also produced wire
coils with a Teflon core since it has a lower loss tangent and a lower
dielectric constant than Rogers 4350B.  Thus, the losses will be less
affected by the core material and the self-resonance frequency of the
coil will be higher. 
Another option to minimize the losses in the coil can be to use a 
superconducting wire to reduce the ohmic losses of the resonator.

\subsection{Spiral Coil}
\label{sec: spiral coil}
\begin{figure}[!htb]
	\begin{center}
		\includegraphics[scale=.3]{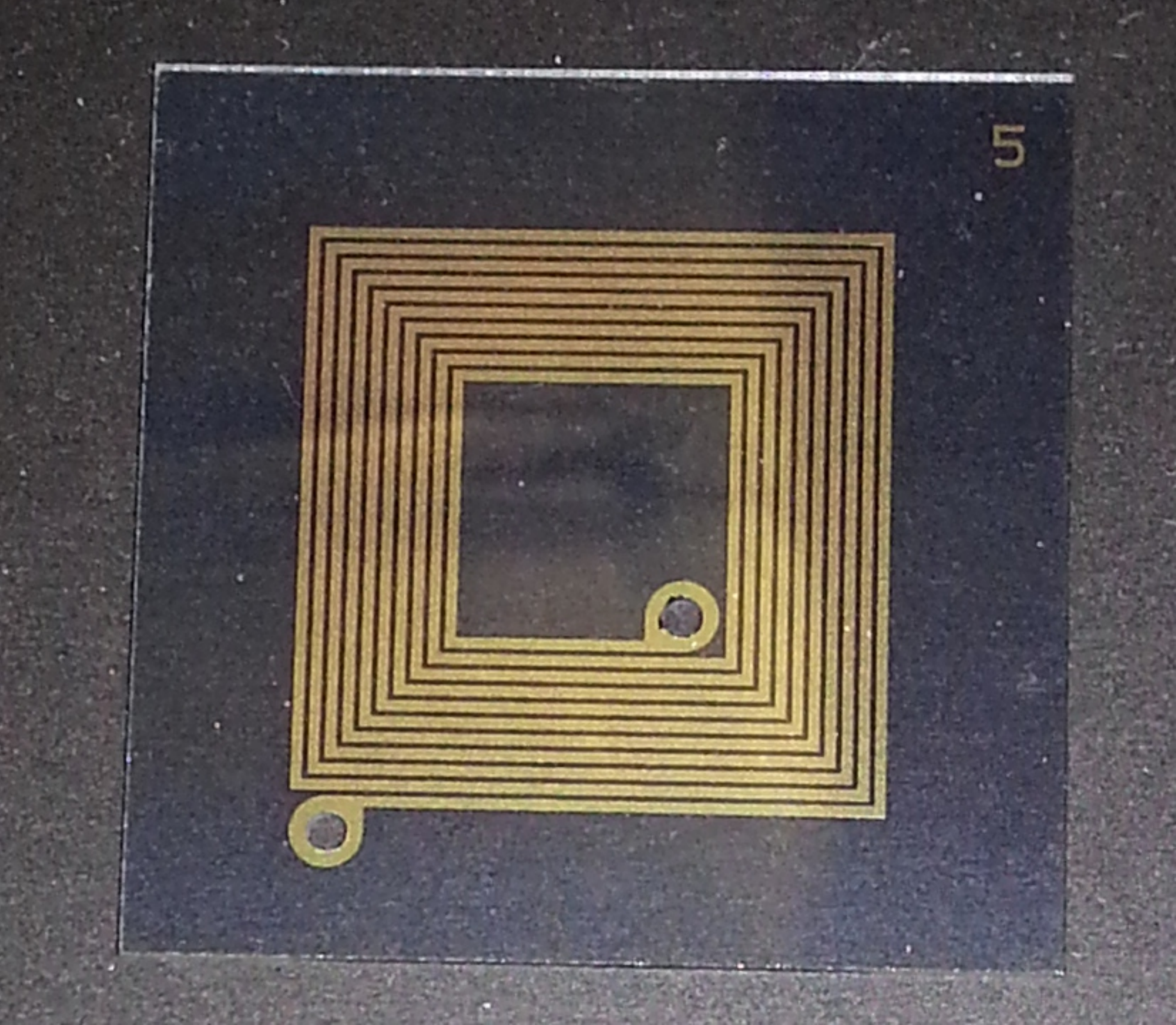}
		\caption{An yttrium barium copper oxide (YBCO) coil with a 200nm 
			protection layer of gold on a 25mm x 25mm piece of sapphire.}
		\label{fig: HTS coil}
	\end{center}
\end{figure}

The quality factor of the presented coils are limited by the ohmic
resistance of the conductor. Thus, a coil based on a superconductor is
expected to result in lower losses in the resonator. Furthermore, it could
be necessary to operate the experiment at temperatures above 15K to
reduce the liquid helium consumption in wet cryostats. At these temperatures, 
a superconductor with a 
high critical temperature such as a high temperature superconductor (HTS) 
is required. However, a coil made from a readily
available HTS material needs to be a two-dimensional (2D) structure due 
to manufacturing constraints, and we choose to use a spiral coil as shown 
in Fig.~\ref{fig: HTS coil}. \\
Since such coils require only 2D structuring, they can be
manufactured in a very accurate and reproducible way.  For the
demonstrated spiral coil, a high-temperature superconductor (HTS) was
produced by \textit{Ceraco Ceramic Coating}\footnote{Ceraco Ceramic 
Coating GmbH, Rote-Kreuz-Str. 8, D-85737 Ismaning, Germany}.  Our coils
were made out of an yttrium barium copper oxide (YBCO) film of 330nm 
thickness on a sapphire wafer.  To
facilitate soldered connections, a 200nm film of gold was placed on
top of the superconductor.  The critical temperature of this YBCO film
is above 87K, and its critical current density is higher than $2 \cdot
10^6$A/cm$^2$ at 77K.  We can derive the peak current in the coil from
Section~\ref{sec: general considerations} to be 0.64A.  Hence, traces
with a width of more than 100$\mu$m are required to stay below the
critical current density. \\
For our square spiral coils, we chose a trace width of 300$\mu$m, a
trace gap of 150$\mu$m, and 10 windings.  We designed 9 different
coils varying the outer diameter between 14 and 20mm.  Equation 4.1 of
reference~\cite{Ref:2Dcoils} gives an estimate for the inductance, which
results in target inductances between 1.2$\mu$H and 1.6$\mu$H for our coils.
This design should allow for coils with very high quality
factors already at temperatures accessible with liquid nitrogen
instead of liquid helium. \\
It should be noted here that superconductors are perfect diamagnets, 
which are spatial inhomogeneities for magnetic fields. One can align
the 2D plane of the spiral coil with the spatially homogeneous magnetic 
field of the experiment to minimize this effect, but one should 
always keep this in mind.

\section{Impedance Matching}
\label{sec: matching}
The resonator is connected to the source through a cable with a given
wave resistance $R_{\mathrm{wave}}$, typically 50$\Omega$.  The complex voltage
reflection coefficient $\underline{r}_{\mathrm{v}}$  at the transition from the cable to
the resonator~\cite{Ref:Pozar} is then
\begin{equation}
	\underline{r}_{\mathrm{v}} = \frac{\underline{V}_{\mathrm{reflection}}}{\underline{V}_{\mathrm{input}}} = 
		\frac{\underline{Z}_{\mathrm{res}}-R_{\mathrm{wave}}}{\underline{Z}_{\mathrm{res}}+R_{\mathrm{wave}}} \hskip 0.2cm ,
	\label{eq: voltage reflection}
\end{equation}
where $\underline{V}_{\mathrm{reflection}}$ is the complex reflected voltage,
$\underline{V}_{\mathrm{input}}$ is the complex incoming voltage, and
$\underline{Z}_{\mathrm{res}}$ is the complex impedance of the
resonator\footnote{The underline denotes a complex quantity.}.  The
reflected power coefficient $R_{\mathrm{p}}$ is then just
$R_{\mathrm{p}} = \left|\underline{r}_{\mathrm{v}}\right|^2$. Since RLC-series resonators
have a very low impedance on resonance,
$\underline{Z}_{\mathrm{res}} = R_{\mathrm{eff}}$, the transition from a
50$\Omega$-cable to the resonator on resonance will reflect almost all
the inserted power.  In order to minimize the reflection, a properly
designed impedance matching network is
required.
The most commonly used matching network is the $L$-network as it is
the simplest, consisting only of two circuit elements. \\
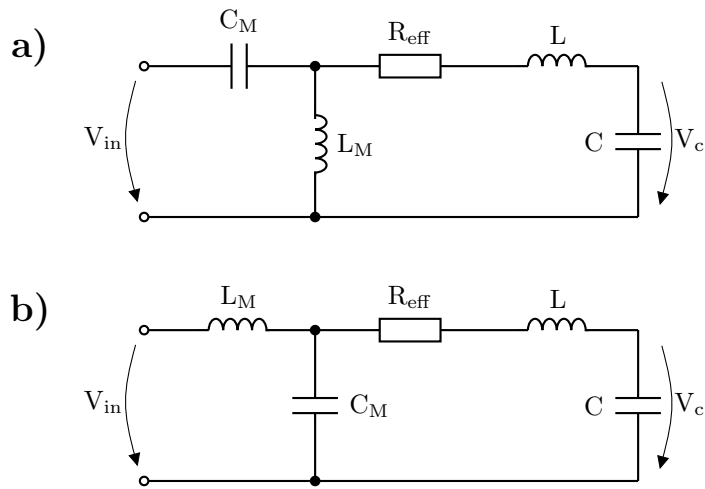
\begin{figure}[ht!]
  \begin{center}
    \begin{circuitikz}[scale=1, transform shape]
      \draw [thick] (0,5.75) to [short, color=black, o-] (1.15,5.75);
			\draw (1.05,5.75) to [/tikz/circuitikz/bipoles/length=1cm,C=$\mathrm{C_M}$] (1.45,5.75);
			\draw [thick] (1.35,5.75) to [short, color=black] (3.1,5.75);
			\draw (3,5.75) to [/tikz/circuitikz/bipoles/length=1cm,R=$\mathrm{R_{eff}}$] (4,5.75);
			\draw [thick] (3.9,5.75) to [short, color=black] (5.05,5.75);
			\draw (4.9,5.75) to [/tikz/circuitikz/bipoles/length=1cm, L, l=$\mathrm{L}$] (6,5.75);
			\draw [thick] (5.8,5.75) to [short, color=black] (6.5,5.75)
				to [short, color=black] (6.5,4.85);
			\draw (6.5,4.95) to [/tikz/circuitikz/bipoles/length=1cm,C, l_=$\mathrm{C}$] (6.5,4.55);
			\draw [thick] (6.5,4.65) to [short, color=black] (6.5,3.75)
				to [short, color=black, -o] (0,3.75);
			\draw (0,3.85) to [open, v^=$\mathrm{V_{in}}$] (0,5.75);
      \draw [thick] (2.25,5.75) to [short, color=black, *-] (2.25,5.05);
			\draw (2.25,5.25) to [/tikz/circuitikz/bipoles/length=1cm, L=$\mathrm{L_M}$, color=black] (2.25,4.15);
      \draw [thick] (2.25,4.35) to [short, color=black, -*] (2.25,3.75);
			\draw (6.7, 3.8) to [open, v=$\mathrm{V_{c}}$] (6.7,5.75);
			\draw (-1.5,6) node[]{\Large{\textbf{a)}}};
			
			\draw [thick] (0,2.25) to [short, color=black, o-] (0.85,2.25);
			\draw (0.7,2.25) to [/tikz/circuitikz/bipoles/length=1cm,L=$\mathrm{L_M}$] (1.8,2.25);
			\draw [thick] (1.6,2.25) to [short, color=black] (3.1,2.25);
			\draw (3,2.25) to [/tikz/circuitikz/bipoles/length=1cm,R=$\mathrm{R_{eff}}$] (4,2.25);
			\draw [thick] (3.9,2.25) to [short, color=black] (5.05,2.25);
			\draw (4.9,2.25) to [/tikz/circuitikz/bipoles/length=1cm,L, l=$\mathrm{L}$] (6,2.25);
			\draw [thick] (5.8,2.25) to [short, color=black] (6.5,2.25)
				to [short, color=black] (6.5,1.35);
			\draw (6.5,1.45) to [/tikz/circuitikz/bipoles/length=1cm,C, l_=$\mathrm{C}$] (6.5,1.05);
			\draw [thick] (6.5,1.15) to [short, color=black] (6.5,0.25)
				to [short, color=black, -o] (0,0.25);
			\draw (0,0.35) to [open, v^=$\mathrm{V_{in}}$] (0,2.25);
      \draw [thick] (2.25,2.25) to [short, color=black, *-] (2.25,1.35);
			\draw (2.25,1.45) to [/tikz/circuitikz/bipoles/length=1cm,C=$\mathrm{C_M}$, color=black] (2.25,1.05);
      \draw [thick] (2.25,1.15) to [short, color=black, -*] (2.25,0.25);
			\draw (6.7, 0.3) to [open, v=$\mathrm{V_{c}}$] (6.7,2.25);
			\draw (-1.5,2.5) node[]{\Large{\textbf{b)}}};
		\end{circuitikz}
    \caption{Two possible matching networks. In a) the matching capacitor $C_{\mathrm{M}}$ is in series and the matching
			coil $L_{\mathrm{M}}$ parallel to the RLC-series resonator. In b) the matching coil $L_{\mathrm{M}}$ is in series and 
			the matching capacitor $C_{\mathrm{M}}$ parallel to the RLC-series resonator.}
		\label{fig: matching networks}
  \end{center}
\end{figure}
Fig.~\ref{fig: matching networks} shows the two possible
$L$-networks where the reactance parallel to the resonator transforms the
real part of the impedance to be $R_{\mathrm{wave}}$.  The resulting imaginary term
is small compared to $R_{\mathrm{wave}}$ (for low loss resonators) and can be 
compensated with the series reactance. \\
\begin{figure}[!htb]
	\begin{center}
		\includegraphics[scale=.5]{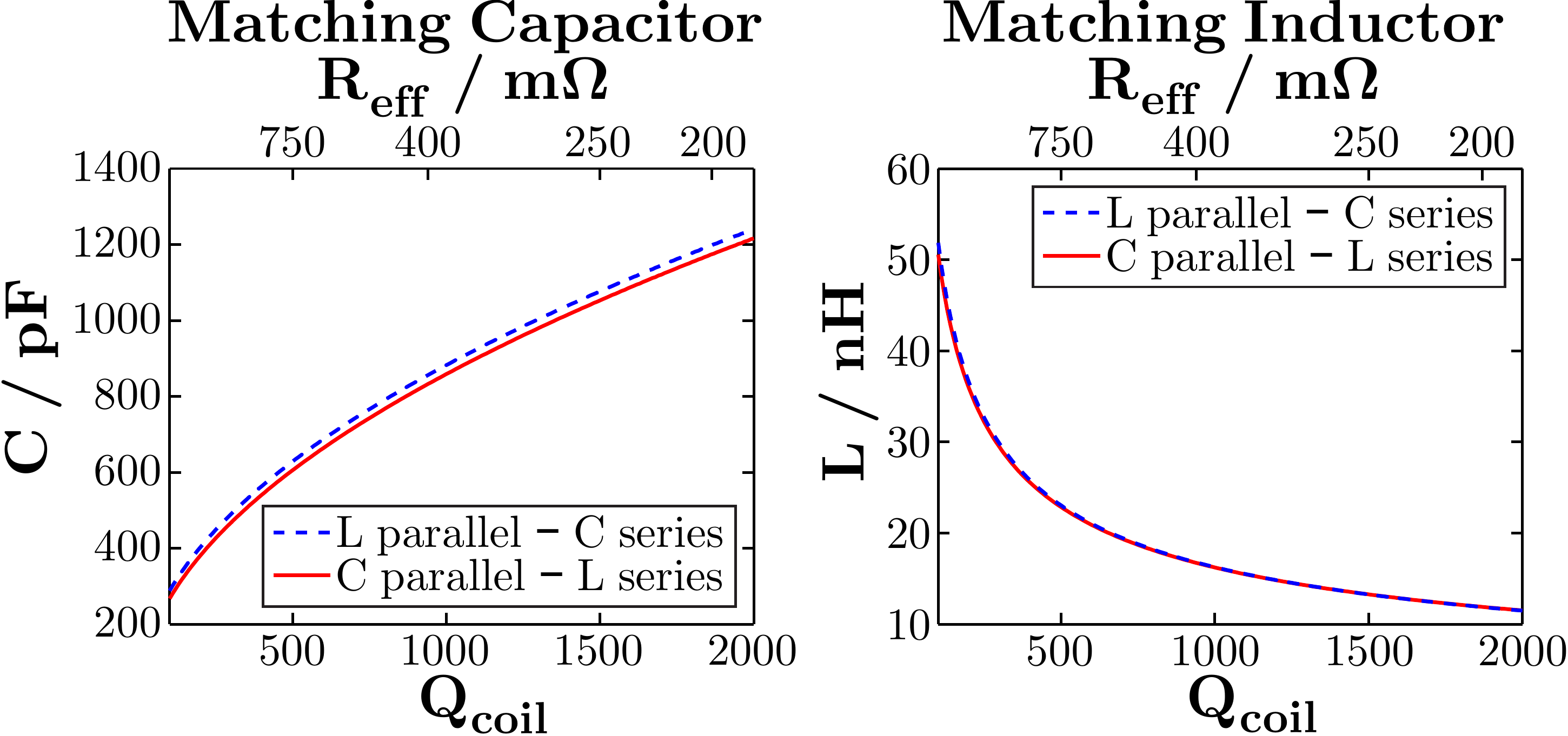}
		\caption{Calculated values of the matching capacitor $C_{\mathrm{M}}$ and matching coil $L_{\mathrm{M}}$ for the matching
			networks of Fig.~\ref{fig: matching networks} over a varying quality factor of the coil. The matching
			is calculated for a 50$\Omega$-cable, an inductor of 1.4$\mu$H, and a capacitor of 10pF.}
		\label{fig: matching network simulation}
	\end{center}
\end{figure}
The resonator's quality factor from~(\ref{eq: quality factor - bandwidth}) can be written as
\begin{equation}
	Q_{\mathrm{coil}} = \frac{\Omega L}{R_{\mathrm{eff}}}
	\label{eq: quality factor - RLC}
\end{equation}
for an RLC-series resonator.  Fig.~\ref{fig: matching network simulation} 
shows the required values for the matching capacitor $C_{\mathrm{M}}$ and inductance 
$L_{\mathrm{M}}$ as a function of the quality factor of the coil with inductance of 
1.4$\mu$H.  For quality factors around 1000, the
required inductance $L_{\mathrm{M}}$ is in the range of 10nH.  Thin traces on a
PCB have approximately a length-dependent inductance of
1nH/mm.  Therefore, small deviations in the trace length will lead to
inefficient matching.  The strong dependence of the matching on the
reactance parallel to the resonator makes the matching network from
Fig.~\ref{fig: matching networks}b) more favorable, because
capacitances of several hundred pF can be adjusted with much more
precision than inductances of about 10nH.  One can incorporate the
inductance $L_{\mathrm{M}}$ into the trace on the PCB from the coaxial connector
to the matching capacitance, by setting this distance between 1 and
2cm.  Hence, we expect a power reflection of 1\% or less even without 
adding an additional inductor. \\
\begin{figure}[ht!]
  \begin{center}
    \begin{circuitikz}[scale=1, transform shape]
			\draw [thick] (-0.5,2.25) to [short, color=black, o-] (3.6,2.25);
			\draw (3.5,2.25) to [/tikz/circuitikz/bipoles/length=1cm,R=$\mathrm{R_{eff}}$] (4.5,2.25);
			\draw [thick] (4.4,2.25) to [short, color=black] (5.45,2.25);
			\draw (5.4,2.25) to [L, l=$\mathrm{L}$] (6.6,2.25);
			\draw [thick] (6.5,2.25) to [short, color=black] (7,2.25)
				 to [short, color=black] (7,1.35);
			\draw (7,1.45) to [/tikz/circuitikz/bipoles/length=1cm,C, l_=$\mathrm{C}$] (7,1.05);
			\draw [thick] (7,1.15) to [short, color=black] (7,0.25)
				to [short, color=black, -o] (-0.5,0.25);
			\draw (-0.5,0.25) to [open, v^=$\mathrm{V_{in}}$] (-0.5,2.25);
			\draw [thick] (1.75,2.25) to [short, color=black, *-] (1.75, 1.35);
			\draw (1.75,1.45) to [/tikz/circuitikz/bipoles/length=1cm,C, l_=$\mathrm{C_M}$, color=black] (1.75,1.05);
			\draw [thick] (1.75,1.15) to [short, color=black, -*] (1.75, 0.25);
			\draw [thick] (2.75,2.25) to [short, color=black, *-] (2.75, 1.65);
			\draw (2.75,1.85) to [/tikz/circuitikz/bipoles/length=1cm,L=$\mathrm{L_{choke}}$, color=black] (2.75,0.65);
			\draw [thick] (2.75,0.85) to [short, color=black, -*] (2.75, 0.25);
			\draw (7.2, 0.3) to [open, v=$\mathrm{V_{c}}$] (7.2,2.25);
		\end{circuitikz}
    \caption{Chosen matching network with only a matching capacitor $C_M$ parallel to the RLC-series resonator. 
			Additionally, one can put an RF-choke parallel to $C_M$ to avoid a possible charging up of $C$. }
		\label{fig: matching network choice}
  \end{center}
\end{figure}
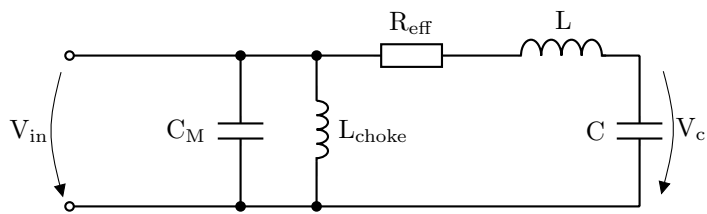
RF transformers or baluns on the RF input side are commonly used to
avoid ground loops between the RF source and the experiment.  When
using the circuit from Fig.~\ref{fig: matching networks}b) in
combination with a transformer, the DC potential of the trap
electrodes is not defined.  We mitigate this by adding an RF-choke
parallel to the matching capacitor, where $\Omega L_{\mathrm{choke}} \gg
\frac{1}{\Omega C_{\mathrm{M}}}$, as depicted in 
Fig.~\ref{fig: matching network choice}. \\
\begin{figure}[!htb]
	\begin{center}
		\includegraphics[scale=.5]{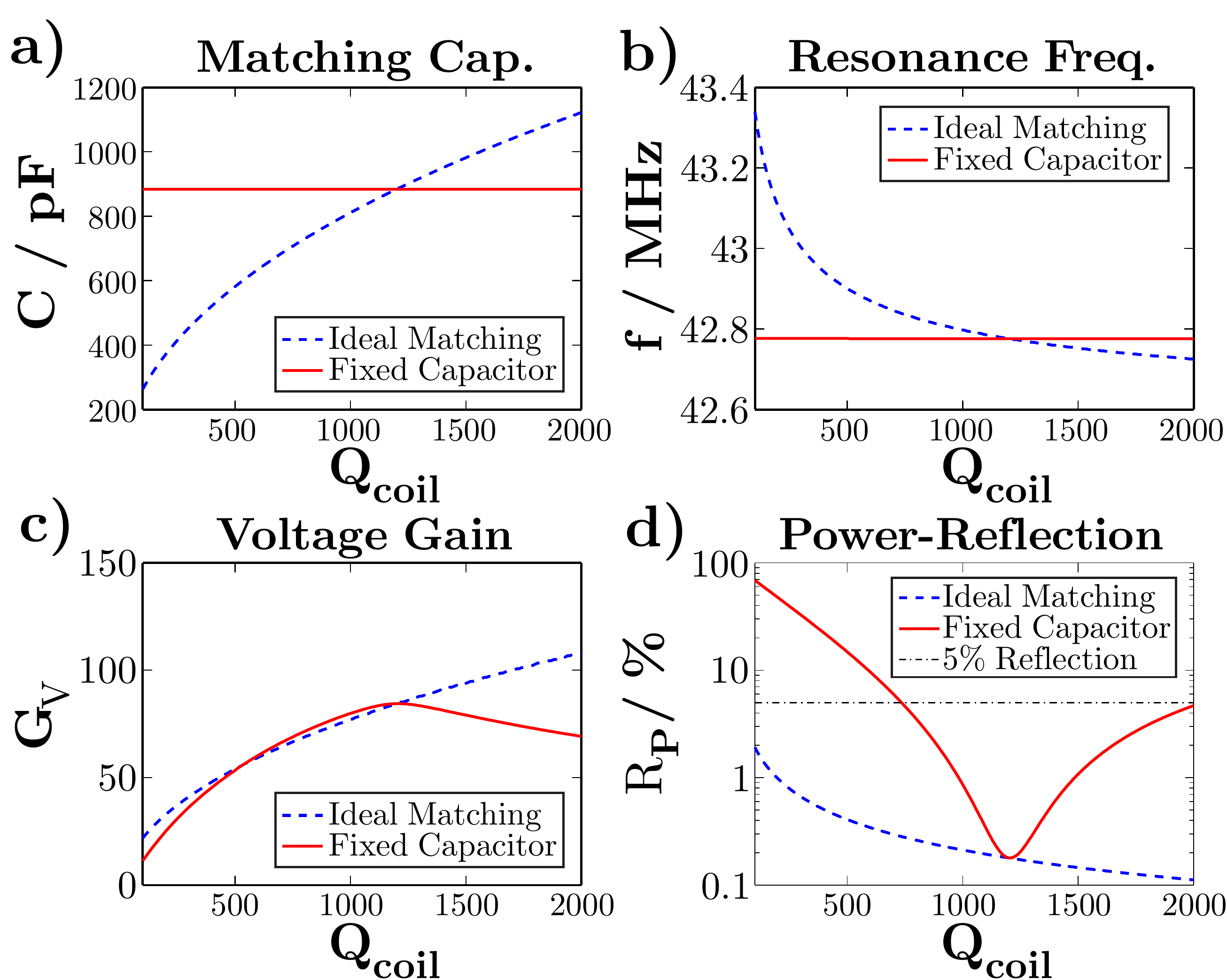}
		\caption{The matching of these plots is calculated for a 50$\Omega$-cable, an inductor of 1.4$\mu$H, and a 
			capacitor of 10pF, thus a resonance frequency of 42.6MHz. The red lines show the values for a fixed capacitance
			$C_{\mathrm{M}}$ chosen for an ideal matching of $Q_{\mathrm{coil}} = 1200$. 
			In a), one can see the required matching capacitor as a function of the quality factor of the coil $Q_{\mathrm{coil}}$. 
			b) shows the change in resonance frequency of an ideal match. In c), the voltage gain is plotted over $Q_{\mathrm{coil}}$. 
			d) shows the reflection coefficient of the matching network. }
		\label{fig: matching reflection}
	\end{center}
\end{figure}
In a cryostat, it is impractical to adjust the matching capacitance
$C_{\mathrm{M}}$ during or after a cool down.  Hence, we have to find an
efficient way of choosing the best value for $C_{\mathrm{M}}$ with a minimum
number of temperature cycles.  For this, we simulated our matching
circuit for varying and also for a constant value of
$C_{\mathrm{M}}$. Fig.~\ref{fig: matching reflection} shows results of
simulations where the fixed capacitor value was chosen for a match at
$Q_{\mathrm{coil}} = 1200$.  Fig.~\ref{fig: matching reflection}a) depicts an
increase of the optimum $C_{\mathrm{M}}$ with increasing
$Q_{\mathrm{coil}}$. Fig.~\ref{fig: matching reflection}b) illustrates the
dependence of the matched resonance frequency on $Q_{\mathrm{coil}}$.  In
Fig.~\ref{fig: matching reflection}c), the simulations confirm the
dependence of the voltage gain on $Q_{\mathrm{coil}}$, following from~(\ref{eq:
  matching 2}).  If the physical value for $Q_{\mathrm{coil}}$ is smaller than
the expected value, used to chose the fixed matching capacitor, the
voltage gain is about the same than for an ideal match.  But if the
actual quality factor $Q_{\mathrm{coil}}$ is higher than the expected value,
the voltage gain will be lower than with an ideal match.  Hence
underestimating the quality factor of the coil should be avoided since
it would result in a significantly lower voltage gain. \\
Fig.~\ref{fig: matching reflection}d) shows the reflection coefficient
as a function of $Q_{\mathrm{coil}}$.  The effect of $L_{\mathrm{M}}$ is small and thus it
was omitted for these simulations.  We can see that even without
$L_{\mathrm{M}}$, but with an ideally matched $C_{\mathrm{M}}$, a low reflection coefficient
can be achieved.  If one chooses a constant value for $C_{\mathrm{M}}$ at a match
for $Q_{\mathrm{coil}} = 1200$, the reflected power will stay
below 5\% even if $Q_{\mathrm{coil}}$ varies between 800 and 2000. \\
The parameters for exact impedance matching can be estimated with one
additional cooling cycle. For the first cycle, one has to guess the
value for $Q_{\mathrm{coil}}$ and adjust $C_{\mathrm{M}}$ according to simulations
following the approach of Fig.~\ref{fig: matching reflection}a).  When
the resonator is cold, the reflection coefficient is measured and
compared with the simulations in Fig.~\ref{fig: matching
  reflection}d).  The measured reflection coefficient can then only
correspond to two values for $Q_{\mathrm{coil}}$.  If the reflection
coefficient passed a minimum during cool-down, one should choose the
higher value, if not, the lower value of $Q_{\mathrm{coil}}$. With this new
value for $Q_{\mathrm{coil}}$, one can calculate the correct matching capacitor
value $C_{\mathrm{M}}$.  Our experience shows that this method will yield a power
reflection coefficient of less than 2\% for the second cool down.

\section{RF Shield}
\label{sec: rf shield}
\begin{figure}[!htb]
	\begin{center}
		\includegraphics[scale=.3]{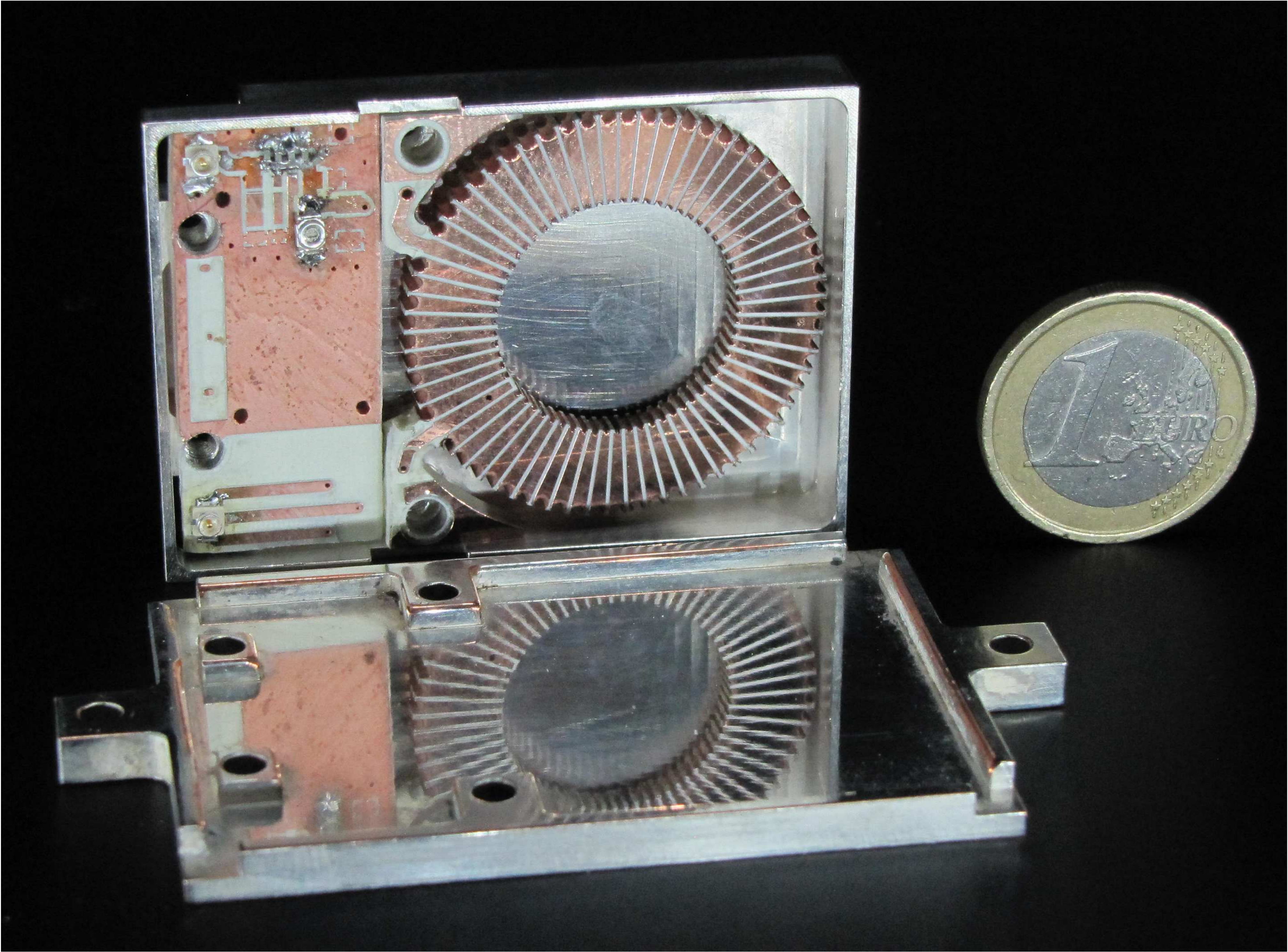}
		\caption{The RLC-resonator built with a PCB coil (right) and the matching network (left) in a silver-plated copper shield.
			The dimensions of the shielded resonator are 57mm x 40mm x 10.2mm.}
		\label{fig: resonator with shield}
	\end{center}
\end{figure}

RF resonators emit electromagnetic radiation, which needs to be
shielded to minimize unwanted RF interference in other parts of the
experiment.  The RF shield needs to be grounded, and thereby generates a
capacitance between the resonator and ground.  This capacitance has to
be minimized as suggested in~(\ref{eq: matching 2}).  In order to keep
the resonator compact, we have to find a trade-off between the size of the
shield and the capacitance added by it. \\
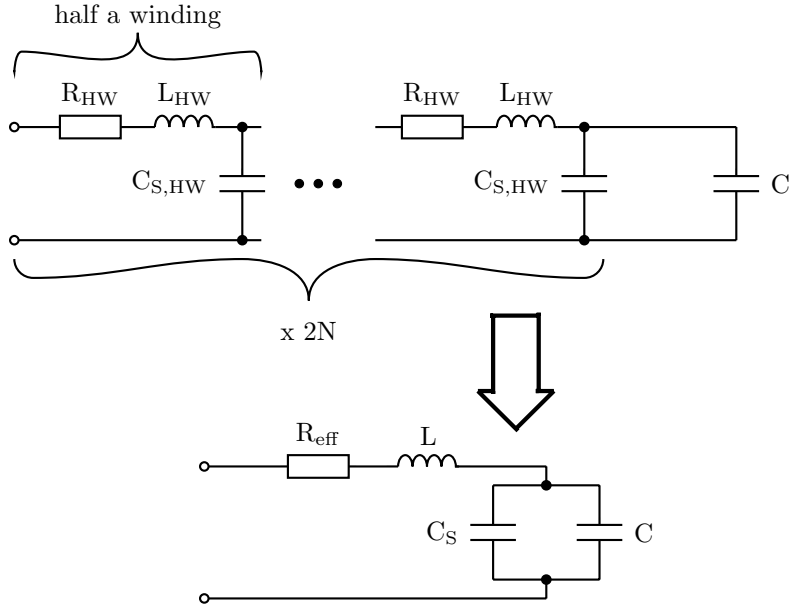
\begin{figure}[ht!]
  \begin{center}
    \begin{circuitikz}[scale=1, transform shape]
			\draw [thick] (1,2.5) to [short, color=black, o-] (2.1,2.5);
			\draw (2,2.5) to [/tikz/circuitikz/bipoles/length=1cm,R=$\mathrm{R_{eff}}$] (3,2.5);
			\draw [thick] (2.9,2.5) to [short, color=black] (3.55,2.5);
			\draw (3.4,2.5) to [/tikz/circuitikz/bipoles/length=1cm, L=$\mathrm{L}$] (4.5,2.5);
			\draw [thick] (4.3,2.5) to [short, color=black] (5.5,2.5)
				to [short, color=black] (5.5,2.25)
				to [short, color=black] (6.2,2.25)
				to [short, color=black] (6.2,1.735);
			\draw (6.2,1.825) to [/tikz/circuitikz/bipoles/length=1cm,C=$\mathrm{C}$] (6.2,1.425);
			\draw [thick] (6.2,1.525) to [short, color=black] (6.2,1)
				to [short, color=black] (5.5,1)
				to [short, color=black] (5.5,0.75)
				to [short, color=black, -o] (1,0.75);
			\draw [thick] (5.5,2.25)	to [short, color=black, *-] (4.8,2.25)
				to [short, color=black] (4.8,1.725);
			\draw (4.8,1.825) to [/tikz/circuitikz/bipoles/length=1cm,C, l_=$\mathrm{C_S}$] (4.8,1.425);
			\draw [thick] (4.8,1.525) to [short, color=black] (4.8,1)
				to [short, color=black, -*] (5.5,1);
			\draw [ultra thick] (5.1,3) to [short, color=black] (5.6,3.5)
				to [short, color=black] (5.362,3.5)
				to [short, color=black] (5.4,3.5)
				to [short, color=black] (5.4,4.538)
				to [short, color=black] (5.4,4.5)
				to [short, color=black] (4.762,4.5)
				to [short, color=black] (4.8,4.5)
				to [short, color=black] (4.8,3.462)
				to [short, color=black] (4.8,3.5)
				to [short, color=black] (4.6,3.5)
				to [short, color=black] (5.127,2.973);
			\draw [thick] (-1.5,7) to [short, color=black, o-] (-0.9,7);
			\draw (-1,7) to [/tikz/circuitikz/bipoles/length=1cm, R=$\mathrm{R_{HW}}$] (0,7);
			\draw [thick] (-0.1,7) to [short, color=black] (0.35,7);
			\draw (0.25,7) to [/tikz/circuitikz/bipoles/length=1cm, L=$\mathrm{L_{HW}}$] (1.25,7);
			\draw [thick] (1.15,7) to [short, color=black] (1.75,7);
			\draw [thick] (1.5,7) to [short, color=black, *-] (1.5,6.35);
			\draw (1.5,6.45) to [/tikz/circuitikz/bipoles/length=1cm, C, l_=$\mathrm{C_{S,HW}}$] (1.5,6.05);
			\draw [thick] (1.5,6.15) to [short, color=black, -*] (1.5,5.5);
			\draw [thick] (1.75,5.5) to [short, color=black, -o] (-1.5,5.5);
			
			\draw [thin] (2.24,6.25) to [short, color=black, -*] (2.25,6.25);
			\draw [thin] (2.51,6.25) to [short, color=black, -*] (2.5,6.25);
			\draw [thin] (2.74,6.25) to [short, color=black, -*] (2.75,6.25);
			
			\draw [thick] (3.25,7) to [short, color=black] (3.6,7);
			\draw (3.5,7) to [/tikz/circuitikz/bipoles/length=1cm, R=$\mathrm{R_{HW}}$] (4.5,7);
			\draw [thick] (4.4,7) to [short, color=black] (4.85,7);
			\draw (4.75,7) to [/tikz/circuitikz/bipoles/length=1cm, L=$\mathrm{L_{HW}}$] (5.75,7);
			\draw [thick] (5.65,7) to [short, color=black] (8,7)
				to [short, color=black] (8,6.35);
			\draw [thick] (6,7) to [short, color=black, *-] (6,6.35);
			\draw (6,6.45) to [/tikz/circuitikz/bipoles/length=1cm, C, l_=$\mathrm{C_{S,HW}}$] (6,6.05);
			\draw [thick] (6,6.15) to [short, color=black, -*] (6,5.5);
			\draw (8,6.45) to [/tikz/circuitikz/bipoles/length=1cm, C=$\mathrm{C}$] (8,6.05);
			\draw [thick] (8,6.15) to [short, color=black] (8,5.5)
				to [short, color=black] (3.25,5.5);
			\draw[>=latex,-,color=black,text=black, thick] (-1.5,7.75) 
				to[out=-90,in=-180] (-1.4,8) to[out=0,in=-90] (0.125,8.25)
				to[out=-90,in=-180] (1.65,8) to[out=-0,in=-90] (1.75,7.75);
			\filldraw[fill=black] (0.125,8.75) circle(0pt)node[anchor=north]{half a winding};
			\draw[>=latex,-,color=black,text=black, thick] (-1.5,5.25) 
				to[out=-90,in=180] (-1.25,5) to[out=0,in=180] (1.5,5.25) to[out=0,in=+90] (2.375,4.7)
				to[out=+90,in=+180] (3.25,5.25) to[out=0,in=180] (6,5) to[out=0,in=-90] (6.25,5.25);
			\filldraw[fill=black] (2.375,4.05) circle(0pt)node[anchor=south]{x 2N};
		\end{circuitikz}
    \caption{Model for the capacitance of the RF shield. From the resistance $R_{\mathrm{HW}}$, the
			inductance $L_{\mathrm{HW}}$, and the capacitance to the shield $C_{\mathrm{S,HW}}$ of half a winding, one
			can calculate the total impedance of the circuit, which allows one to
			derive the total capacitance $C_{\mathrm{S}}$ caused by the shield.}
		\label{fig: shield cap model}
  \end{center}
\end{figure}
First, we need to find a model for the shield capacitance
$C_{\mathrm{S}}$. Fig.~\ref{fig: shield cap model} depicts a model where we
divide the coil in 2$N$ half windings. Each has an ohmic resistance of
$R_{\mathrm{HW}}$, an inductance $L_{\mathrm{HW}}=L/2N$, and a capacitance to the shield
$C_{\mathrm{S,HW}}$.  We can estimate $C_{\mathrm{S,HW}}$ with the standard formula for
a parallel-plate capacitor if we assume that the grounded shield is
parallel to the coil.  From the calculations of the impedance of the
entire circuit, we can extract the effective capacitance $C_{\mathrm{S}}$ that
occurs due to the presence of the shield.
\\
\begin{figure}[!htb]
	\begin{center}
		\includegraphics[scale=.5]{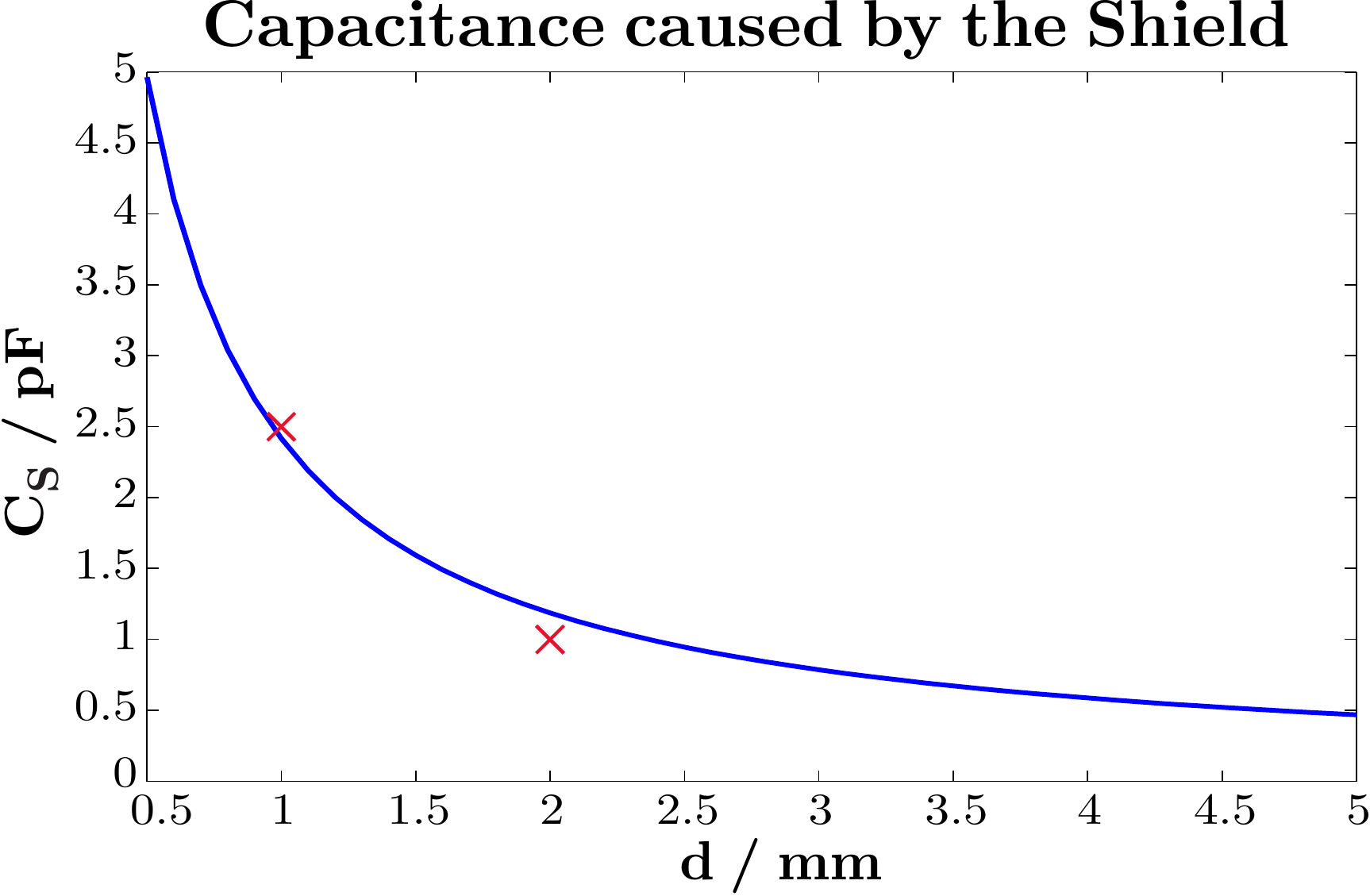}
		\caption{Simulated capacitance caused by the shield in
                  dependence of the distance between the coil and the
                  shield. The red crosses are measured capacitances
                  for two different shield-coil distances.}
		\label{fig: shield cap simulation}
	\end{center}
\end{figure}
The results of these simulations for our PCB coils with an outer
radius of 17.5mm and an inner radius of 10.2mm are shown in 
Fig.~\ref{fig: shield cap simulation}.  It can be seen that a distance of
about 2mm is sufficient to keep the additional capacitive load below
1pF.  Adapted simulations for the wire coils yield results similar to
the ones for the PCB coils. Hence, we can use the same shield for both
types of coils. \\
We expect that the capacitance caused by the shield for the spiral
coils is even smaller, because its surface is smaller.  Additionally, if 
we place the spiral coil in the center of the same shield, the distance 
to the shield will be 1.55mm bigger than for the other coils. 
Hence, the capacitance to ground will decrease further.
Moreover, the shield has an influence on the magnetic field lines of
the spiral coil in this configuration and thereby also on the inductance
and the losses of the coil. \\
At room temperature, the penetration depth due to skin effect in
copper at frequencies around 40MHz is about 10$\mu$m.  Thus, any
mechanically stable shield, of a couple of 1/10mm thickness, 
should yield a suitable attenuation.
Fig.~\ref{fig: resonator with shield} shows a resonator encased in a
silver-plated copper shield.  The dimensions of this resonator are 57mm
x 40mm x 10.2mm.  On the left side of the resonator, one can see a PCB
with the matching network at the top and the SMT-connector for the
inductive pick-up at the bottom. On the right side is the PCB coil,
and the two PCBs are connected together by soldered joints.

\section{Results}
\label{sec: results}
In our measurement setup, we used a capacitive pick-up to measure the
voltage at the trap.  With the known ratio of this capacitive voltage
divider, we could directly measure the voltage at the trap and
calibrate the inductive pick-up of the coil. \\
\begin{table}[hbtp]
	\caption{Measured resonance frequencies, voltage gains, and quality factors of the tested coils.}
	\label{tab: measurement results}
	\centering
	\begin{tabular}{lcccc}
		\hline\noalign{\smallskip}
		\textbf{Name} & \textbf{T} & $\mathbf{f_{res}}$ & $\mathbf{G_V}$ & $\mathbf{Q_{coil}}$  \\
		\noalign{\smallskip}\hline\noalign{\smallskip}
		PCB coil & 295K & 42.4MHz & 19.4 & 60 \\
		PCB coil & $\sim$80K & 42.8MHz & 27.1 & 115 \\
		PCB coil & $\sim$10K & 42.9MHz & 41.5 & 208 \\
		\noalign{\smallskip}\hline
		Wire Coil & 295K & 36.8MHz & 25.2 & 89 \\
		Wire Coil & $\sim$80K & 37.2MHz & 36.5 & 184 \\
		Wire Coil & $\sim$10K & 37.9MHz & 58.9 & 624 \\
		\noalign{\smallskip}\hline
		HTS Coil & $<$88K & 44.3MHz & 92.1 & 1172 \\
		\noalign{\smallskip}\hline
	\end{tabular}
\end{table}
Table~\ref{tab: measurement results} shows our measurement
results.  The PCB coils had an inductance very close to the desired
value.  Unfortunately, we could not achieve the expected quality
factor.  We had designed the wire coils to have losses
similar to the PCB coils, but the PCB coils showed much higher losses
than the wire coils.  We think this is due to an imperfect production
process that we could not improve further with our in-house techniques.\\
Additionally to the PCB coils in copper, we produced them in silver as 
well.  The inductances were similar, but the losses in the silver coils 
were even higher.  This was unexpected as silver is a better conductor 
than copper.  But the silver coils needed to be reworked after fabrication
to remove shorts between the segments.  During this process, the silver coating
might have been scratched, resulting in higher losses.  We expect that
silver plating after structuring should reduce the losses compared to
a copper coil, but we were unable to do so with our in-house
fabrication process. \\
The voltage gain of about 60 of the wire coil would require 160mW RF
input power to get to the desired 170$V_{\mathrm{rms}}$ at the trap.  If we
adjust the voltage to keep the stability parameter $q = 0.25$, only
100mW RF input power are required at the modified resonance frequency. \\
We expect that one can improve the quality factor of the wire coil by
using a thicker wire.  We also tested a resonator with a lead-plated
wire, which did not show an increased voltage gain.  We either could
not reach the critical temperature of lead in our test setup, or the
lead was not pure enough to become superconducting. \\
The spiral coil showed the highest voltage gain, although the shield
influences the magnetic field generated by the coil.  We observed a
change of the resonance frequency of a couple of percent when cooling
down from liquid nitrogen temperatures to liquid helium temperatures,
which we attribute to the shield surrounding the coil.  With a
voltage gain of 92, one would need about 70mW of RF input power to get
170$V_{\mathrm{rms}}$.  With a stability parameter $q=0.25$ at the higher
resonance frequency, it would be 82mW.

\section{Conclusion}
\label{sec: conclusion}
We have demonstrated three designs of RF resonators for trapped ion
experiments with Paul traps.  The PCB coils have favorable mechanical
properties but still need an improved production process, such as
e.g. thermal annealing.  The wire coils fulfill our criteria and can
be further improved by using a thicker or superconducting wire.  The
HTS spiral coil showed the highest voltage gain, which was already 
accessible at temperatures reached with liquid nitrogen cooling. \\
Efficient matching is usually difficult in a cryogenic environment
because normally one cannot tune the matching parameters when the
cryostat is cold.  We have shown a recipe to efficiently match
RLC-resonators without precise knowledge of their quality factors.
During this empirical study we regularly achieved matching with a
power reflection of less than 2\% on the second cool-down.  Additionally, we
designed and implemented a shield for resonators to minimize RF
pick-up in other parts of the experiment.

\section{Acknowledgments}
\label{sec: acknowledgments}
We thank Gerhard Hendl from the Institute of Quantum Optics and Quantum 
Information of the Austrian Academy of Science for the production of PCBs, 
and PCB coils for this study. Furthermore, we want to thank Florian Ong for 
feedback on the manuscript.
This research was funded by the Office of the Director of National Intelligence (ODNI),
Intelligence Advanced Research Projects Activity (IARPA), through the Army Research Office grant
W911NF-10-1-0284. All statements of fact, opinion or conclusions contained herein are those of the
authors and should not be construed as representing the official views or policies of IARPA, the ODNI,
or the U.S. Government.

\end{document}